\documentclass[a4paper,11pt]{article}
\pdfoutput=1 

\usepackage{jcappub} 

\usepackage[T1]{fontenc} 
\usepackage{fancyvrb}
\usepackage{fancybox}

\usepackage{dcolumn}
\usepackage{bm}
\usepackage{color}
\usepackage{amsmath}
\usepackage{amssymb}

\usepackage{tabularray}
\usepackage{multirow}

\usepackage{mathcomp}
\usepackage[colorlinks=true,linkcolor=blue,citecolor=blue,urlcolor=blue]{hyperref}

\usepackage[hang,small,bf]{caption}
\usepackage[subrefformat=parens]{subcaption}
\thisfancyput(12.0cm,0cm){{KUNS-3125}, {YITP-26-101}}

\title{
Constraining Primordial Power Asymmetry from Galaxy Clustering and Peculiar Velocity Information
}

\author[a]{Keita Minato,}
\author[b,c,d]{Atsushi Taruya,}
\author[e,c]{Teppei Okumura,}
\author[f]{and Maresuke Shiraishi}

\affiliation[a]{Department of Physics, Kyoto University,\\
Kyoto 606-8502, Japan}

\affiliation[b]{Center for Gravitational Physics and Quantum Information,\\
Yukawa Institute for Theoretical Physics, Kyoto University,\\
Kyoto 606-8502, Japan}

\affiliation[c]{Kavli Institute for the Physics and Mathematics of the Universe (WPI),\\
The University of Tokyo Institutes for Advanced Study, The University of Tokyo,\\
5-1-5 Kashiwanoha, Kashiwa, Chiba 277-8583, Japan}

\affiliation[d]{Korea Institute for Advanced Study,\\
85 Hoegiro, Dongdaemun-gu, Seoul 02455, Republic of Korea}

\affiliation[e]{Academia Sinica Institute of Astronomy and Astrophysics (ASIAA),\\
No. 1, Section 4, Roosevelt Road, Taipei 106319, Taiwan}

\affiliation[f]{Department of Economics, Management and Information Science,\\
Onomichi City University, Onomichi, Hiroshima 722-8506, Japan}

\emailAdd{minato@tap.scphys.kyoto-u.ac.jp}
\emailAdd{ataruya@yukawa.kyoto-u.ac.jp}
\emailAdd{tokumura@asiaa.sinica.edu.tw}
\emailAdd{shiraishi@onomichi-u.ac.jp}

\abstract{

{Primordial power asymmetry would provide a distinctive probe of departures from statistical isotropy, offering an important clue to the physics of the primordial Universe. We investigate how well such asymmetries can be tested with galaxy surveys and
quantify how peculiar-velocity information can improve these tests. We develop a unified bipolar spherical harmonic (BipoSH) analysis of the auto- and cross-power spectra of galaxy density and line-of-sight peculiar velocity, and perform Fisher forecasts for a joint analysis of a Euclid-like spectroscopic galaxy survey and peculiar velocities reconstructed from Simons Observatory-like CMB maps through the kinetic Sunyaev--Zel'dovich effect.}

{For dipolar asymmetry, we consider both scale-independent and scale-dependent modulations proportional to $k^{-0.5}$. Galaxy clustering provides most of the constraining power in both cases, while peculiar velocities add only modest information. For quadrupolar asymmetry, we account for the recently identified anisotropic galaxy-bias response and marginalize over its amplitude. We find that galaxy clustering alone suffers from a degeneracy between the primordial quadrupolar modulation and anisotropic galaxy bias. This degeneracy can be substantially broken by adding peculiar-velocity information: the density--velocity cross-spectrum provides complementary information to the galaxy auto-spectrum, while the velocity auto-spectrum constrains the primordial modulation independently of anisotropic galaxy bias. The velocity information becomes increasingly effective for more negative scale dependence, yielding constraints tighter than those from BOSS
{
for scale dependences proportional to $k^{-1}$ and $k^{-2}$.
}
Our results demonstrate that peculiar-velocity information provides a complementary avenue for testing primordial power asymmetry with large-scale structure, particularly for quadrupolar asymmetry in the presence of anisotropic galaxy bias.
}
}
\begin{document}

\maketitle
\flushbottom

\section{Introduction}
\label{sec:intro}

The standard cosmological model is based on the cosmological principle,
namely that the Universe is statistically homogeneous and isotropic on
sufficiently large scales.
This assumption is strongly supported by a wide range of observations
and provides the foundation of modern
cosmology~\cite{Scrimgeour:2012wt,Planck:2019evm}.
Nevertheless, testing isotropy remains important, since any robust
evidence for a preferred direction in primordial fluctuations would
provide a clue to physics beyond the simplest isotropic scenarios of
the early Universe.

Large-scale structure (LSS) provides a powerful and complementary way
to test primordial power asymmetry.
Compared with the cosmic microwave background (CMB), LSS contains
three-dimensional information and a much larger number of accessible
Fourier modes~\cite{Shiraishi:2016wec,Shiraishi:2020pea}.
It therefore offers an important opportunity to probe directional
dependence of primordial fluctuations with different systematics and
independent observational
data~\cite{Hirata:2009ar,Yoon:2014cba,Appleby:2014kea,
Bengaly:2016qtf,Tiwari:2015tba}.

{Two forms of primordial power asymmetry have received particular
attention. One is}
dipolar power asymmetry, motivated by large-scale anomalies reported
in the cosmic microwave background
(CMB)~\cite{Gordon:2006ag,Erickcek:2008sm,Schmidt:2012ky}.
In this scenario, the fluctuation amplitude is modulated along a
preferred direction, producing a hemispherical difference in power
across the sky.
Analyses of CMB data have reported such a signal on large angular
scales, although its physical origin remains
unclear~\cite{Hoftuft:2009rq,Aiola:2015rqa,Planck:2015igc,
Planck:2019evm,Schwarz:2015cma}.

{The other is}
quadrupolar power asymmetry in the primordial power spectrum, which
arises in classes of anisotropic inflationary models and has been
studied extensively as an observational
target~\cite{Ackerman:2007nb,Yokoyama:2008xw,Watanabe:2009ct,
Watanabe:2010fh,Bartolo:2012sd}.
CMB analyses have found no significant evidence for such asymmetry and
constrain the dimensionless quadrupolar modulation amplitude at the
$10^{-2}$ level~\cite{Kim:2013gka,Ramazanov:2016gjl,
Planck:2019evm}.

Large-scale structure provides an independent low-redshift probe of
the same primordial power asymmetry, based on observables and
systematic uncertainties different from those of the CMB.
Ref.~\cite{Sugiyama:2017ggb} applied a bipolar spherical harmonic analysis to BOSS DR12
galaxies and found no significant evidence
{for}
quadrupolar power asymmetry, with typical $1\sigma$ uncertainties of
$\Delta g_{2M}\sim2\times10^{-2}$.
Their galaxy-clustering model attributed the quadrupolar signal to the
primordial modulation of the matter power spectrum and did not include
an anisotropic response of biased tracers.
Ref.~\cite{Shiraishi:2023zda} showed within the bias-expansion
framework that the tensor characterizing primordial quadrupolar power
asymmetry can couple to the tidal field in the tracer-density relation,
generating an additional contribution proportional to
$g_{2M}b_1^{(2)}$, where $g_{2M}$ and $b_1^{(2)}$ denote the
primordial quadrupole and anisotropic-bias coefficients, respectively.
Because this contribution occupies the same $L=2$ angular sector as
the direct primordial modulation, the two contributions are partially
degenerate in galaxy clustering.
Ref.~\cite{Masaki:2024hzn} directly confirmed this anisotropic
halo-bias response in cosmological $N$-body simulations.
They found that $b_1^{(2)}$ is negative for cluster-scale halos and
becomes more negative with increasing halo mass, establishing it as a
tracer-dependent response that must be calibrated or marginalized
over.

These considerations motivate combining galaxy clustering with an
observable that constrains the primordial power asymmetry without
$b_1^{(2)}$.
{Peculiar velocity provides such a complementary probe.
In linear theory, the peculiar-velocity and galaxy-density fields are
related differently to the underlying matter density field, giving
their power spectra different dependences on model
parameters~\cite{Shiraishi:2020pea}.
Peculiar velocity is also relatively more sensitive to long-wavelength
modes than the galaxy-density field.
This long-wavelength sensitivity makes it particularly relevant to
the reported CMB dipolar signal, which is most prominent on large
angular scales.
Moreover, velocity measurements are subject to observational
systematics distinct from those affecting galaxy
clustering~\cite{Smith:2018bpn}.}

{In this work, we consider both dipolar and quadrupolar power
asymmetries and extend the Fisher analysis of
Ref.~\cite{Shiraishi:2020pea}, which used galaxy-density and
peculiar-velocity correlators.
Within this framework, we quantify how peculiar velocity complements
galaxy clustering in each case.
For the dipolar case, we assess the density--velocity cross-spectrum
and velocity auto-spectrum as consistency checks based on
observational systematics different from those of galaxy clustering.
For the quadrupolar case, we include the anisotropic-bias response and
treat $b_1^{(2)}$ as a nuisance parameter.
We examine how the different $g_{2M}$--$b_1^{(2)}$ degeneracy
directions of the galaxy auto- and density--velocity cross-spectra
improve the joint constraint, and how the velocity auto-spectrum
constrains $g_{2M}$ independently of $b_1^{(2)}$.}

This paper is organized as follows.
In Sec.~\ref{sec:probes}, we describe the observables, namely galaxy
clustering and peculiar velocity.
In Sec.~\ref{sec:models}, we introduce the models of power asymmetry
considered in this work.
In Sec.~\ref{sec:biposh}, we present the BipoSH decomposition of the
power spectra used in the analysis.
In Sec.~\ref{sec:fisher}, we formulate the Fisher forecast.
The results are presented in Sec.~\ref{sec:results}, and
Sec.~\ref{sec:conclusion} is devoted to the conclusion.

\section{Probes}
\label{sec:probes}

In this section, we introduce the two observables considered in this work, namely galaxy clustering
and the line-of-sight peculiar velocity field.
We first define these observables in the statistically isotropic limit.  This separates probe-dependent effects from model-dependent asymmetry.
These provide the baseline probe kernels before model-dependent power asymmetry is introduced in Sec.~\ref{sec:models}.

Throughout this paper, we adopt the local plane-parallel approximation~\cite{Hamilton:1997zq,Shiraishi:2016wec,Shiraishi:2020pea}.
Namely, within each local patch of the survey, the line-of-sight direction is treated as a fixed unit vector $\hat{n}$, and we define
\begin{align}
  \mu \equiv \hat{k}\cdot\hat{n},
\end{align}
where $\hat{k}$ is the unit vector along the wavevector $\bm{k}$.

\subsection{Galaxy clustering}
\label{sec:probes_galaxy}

{
Throughout this work, we model both the galaxy-density and peculiar-velocity fields within linear perturbation theory.  Accordingly, the Fisher forecasts are restricted to wavenumbers for which the linear description is expected to be valid, as specified in Sec.~\ref{subsec:survey_setup}.
}

In the statistically isotropic limit, the observed galaxy fluctuation in redshift space is described by the Kaiser formula~\cite{Kaiser:1987qv,Hamilton:1997zq},
\begin{align}
  \delta_{\rm g}(\bm{k},\hat{n})
  =
  K_{\rm g}(\mu)\,\delta_{\rm m}(\bm{k}),
  \label{eq:delta_g_kernel}
\end{align}
with the galaxy kernel
\begin{align}
  K_{\rm g}(\mu)=b_1+f\mu^2.
  \label{eq:Kg_def}
\end{align}
Here $\delta_{\rm m}$ denotes the matter density fluctuation, $b_1$ is the linear galaxy bias, $f\equiv d\ln D/d\ln a$ is the linear growth rate, and $D$ is the linear growth factor.

This expression shows that galaxy clustering depends both on the matter density field itself and on redshift-space distortions.
The term proportional to $b_1$ represents the biased response of galaxies to the matter fluctuation, while the term proportional to $f\mu^2$ is the Kaiser redshift-space distortion.

\subsection{Peculiar velocity}
\label{sec:probes_velocity}

In addition to galaxy clustering, the peculiar velocity field provides a complementary probe of large-scale structure.
Observationally, peculiar velocities can be accessed through several methods.  Among them, the kinetic Sunyaev--Zel'dovich effect\footnote{\label{fn:ksz_velocity_relation}The kSZ temperature fluctuation is related to the line-of-sight peculiar velocity by $\Delta T_{\rm kSZ}=-(T_0\tau/c)v_\parallel$, where $T_0$ is the mean CMB temperature and $\tau$ is the effective optical depth.  The optical depth is not a response coefficient describing how a tracer field responds to the underlying density or tidal fields and therefore does not carry the anisotropic-bias contribution considered here.  Moreover, since the factor $T_0\tau/c$ cancels between the signal derivatives and the covariance in the Fisher analysis presented below, we use $v_\parallel$, rather than $\Delta T_{\rm kSZ}$, as the probe throughout this paper.} 
offers a promising way to measure the line-of-sight velocity of galaxies and clusters~\cite{Hand:2012ui,Deutsch:2017ybc,Smith:2018bpn,Munchmeyer:2018eey,Giri:2020pkk}.
Recent measurements combining ACT DR6 with DESI DR2 have detected the galaxy--velocity cross-correlation for several spectroscopic tracers and the velocity auto-correlation for luminous red galaxies (LRGs), demonstrating the observational feasibility of the power spectra considered here~\cite{Chaussidon:2026vmn}.
Unlike galaxy number density, peculiar velocity directly traces gravitational motion sourced by the total matter distribution.

In the statistically isotropic limit, the line-of-sight peculiar velocity in Fourier space is given by~\cite{Okumura:2013zva,Sugiyama:2015dsa,Shiraishi:2020pea}
\begin{align}
  v_{\parallel}(\bm{k},\hat{n})
  =
  K_{\rm v}(k,\mu)\,\delta_{\rm m}(\bm{k}),
  \label{eq:v_kernel}
\end{align}
with the velocity kernel
\begin{align}
  K_{\rm v}(k,\mu)=i\,\frac{aHf\mu}{k},
  \label{eq:Kv_def}
\end{align}
where $a$ is the scale factor and $H$ is the Hubble parameter.

A crucial difference from galaxy clustering is that peculiar velocity is not a biased tracer of the density field.  In the model considered here, neither the isotropic nor anisotropic bias parameters enter the velocity observable.
Moreover, because $K_{\rm v}\propto k^{-1}$, it is more sensitive to long-wavelength modes than the density field itself.
This feature makes peculiar velocity particularly useful when power asymmetry is enhanced on large scales.

\subsection{Redshift-space power spectra in a statistically isotropic universe}
\label{sec:probes_isotropic_power}

For ${\rm X,Y}\in\{{\rm g,v}\}$, we define the auto- and cross-power spectra by
\begin{align}
  \left\langle
  \Delta_{\rm X}(\bm{k},\hat{n})
  \Delta_{\rm Y}(\bm{k}',\hat{n})
  \right\rangle
  =
  (2\pi)^3\delta_{\rm D}^{(3)}(\bm{k}+\bm{k}')
  P^{\rm XY}(\bm{k},\hat{n}),
  \label{eq:PXY_def}
\end{align}
where
\begin{align}
  \Delta_{\rm g}\equiv \delta_{\rm g},
  \qquad
  \Delta_{\rm v}\equiv v_{\parallel}.
\end{align}

In the statistically isotropic limit, the matter power spectrum depends only on $k$:
\begin{align}
  \left\langle
  \bar{\delta}_{\rm m}(\bm{k})
  \bar{\delta}_{\rm m}(\bm{k}')
  \right\rangle
  =
  (2\pi)^3\delta_{\rm D}^{(3)}(\bm{k}+\bm{k}')
  \bar{P}_{\rm m}(k).
  \label{eq:Pm_iso}
\end{align}
Here, $\bar{\delta}_{\rm m}$ denotes the statistically isotropic part of the matter density fluctuation.
The observable power spectra then take the unified form
\begin{align}
  \bar{P}^{\rm XY}(\bm{k},\hat{n})
  =
  \,
  K_{\rm X}(k,\mu)\,
  K_{\rm Y}^{\ast}(k,\mu)\,
  \bar{P}_{\rm m}(k).
  \label{eq:PXY_iso_kernel}
\end{align}

Since $\bar{P}^{\rm XY}$ depends on $\hat{k}$ and $\hat{n}$ only through $\mu$, it can be expanded in Legendre polynomials $\mathcal{L}_{\ell}$ as
\begin{align}
  \bar{P}^{\rm XY}(\bm{k},\hat{n})
  =
  \sum_{\ell}
  P_{\ell}^{\rm XY}(k)\,
  \mathcal{L}_{\ell}(\mu).
  \label{eq:PXY_legendre}
\end{align}
{
For the galaxy auto-spectrum, the monopole, quadrupole, and
hexadecapole components are non-vanishing:
}
\begin{align}
  P_0^{\rm gg}(k)
  &=
  \left(
  b_1^2+\frac{2}{3}b_1f+\frac{1}{5}f^2
  \right)\bar{P}_{\rm m}(k),
  \\
  P_2^{\rm gg}(k)
  &=
  \left(
  \frac{4}{3}b_1f+\frac{4}{7}f^2
  \right)\bar{P}_{\rm m}(k),
  \\
  P_4^{\rm gg}(k)
  &=
  \frac{8}{35}f^2\bar{P}_{\rm m}(k).
  \label{eq:Pgg_multipoles}
\end{align}
{
For the density--velocity cross-spectrum, the dipole and octupole
components are non-vanishing:
}
\begin{align}
  P_1^{\rm gv}(k)
  &=
  -i\,\frac{aHf}{k}
  \left(
  b_1+\frac{3}{5}f
  \right)\bar{P}_{\rm m}(k),
  \\
  P_3^{\rm gv}(k)
  &=
  -i\,\frac{2aHf^2}{5k}\bar{P}_{\rm m}(k).
  \label{eq:Pgv_multipoles}
\end{align}
{
For the velocity auto-spectrum, the monopole and quadrupole
components are non-vanishing:
}
\begin{align}
  P_0^{\rm vv}(k)
  &=
  \frac{1}{3}
  \left(
  \frac{aHf}{k}
  \right)^2
  \bar{P}_{\rm m}(k),
  \\
  P_2^{\rm vv}(k)
  &=
  \frac{2}{3}
  \left(
  \frac{aHf}{k}
  \right)^2
  \bar{P}_{\rm m}(k).
  \label{eq:Pvv_multipoles}
\end{align}

These isotropic spectra provide the baseline on top of which dipolar or quadrupolar power asymmetry is introduced in the next section.

\section{Models of Power Asymmetry}
\label{sec:models}

Having defined the baseline probe kernels in the statistically isotropic limit in Sec.~\ref{sec:probes}, we now introduce the models of power asymmetry considered in this work.
The model dependence enters through the 
{
modulated
}
matter power spectrum and, in the quadrupolar case, through the 
correction to the galaxy kernel
{
induced by the power asymmetry.
}
To treat the dipolar and quadrupolar cases as uniformly as possible, we label them by $n$, with
\begin{align}
  n=1 &\quad \text{for the dipolar power asymmetry},
  \nonumber\\
  n=2 &\quad \text{for the quadrupolar power asymmetry}.
  \label{eq:n_def}
\end{align}

We write the modulated matter power spectrum in the unified form
\begin{align}
  P_{\rm m}^{(n)}(k,\hat{p}_n)
  =
  \bar{P}_{\rm m}(k)
  \left[
  1+\sum_M h_{nM}(k)\,Y_{nM}(\hat{p}_n)
  \right],
  \label{eq:Pmn_unified}
\end{align}
where
\begin{align}
  \hat{p}_1=\hat{n},
  \qquad
  \hat{p}_2=\hat{k}.
  \label{eq:pn_def}
\end{align}
Here $\bar{P}_{\rm m}(k)$ is the statistically isotropic matter power spectrum, and $h_{nM}(k)$ denotes the 
{
modulation
}
amplitude including its scale dependence. 
The reality of $P_{\rm m}^{(n)}$ requires
\begin{align}
  h_{nM}^{\ast}(k)=(-1)^M h_{n,-M}(k).
  \label{eq:hnM_reality}
\end{align}

The meaning of the directional argument $\hat{p}_n$ differs between the two cases.
For $n=1$, the angular dependence is phenomenological and is written in terms of the observed line-of-sight direction $\hat{n}$, whereas for $n=2$ it is an asymmetry in Fourier-space with angular argument $\hat{k}$.
The two cases therefore differ not only in multipole order but also in the argument of the modulation.

For the dipolar case, we define
\begin{align}
  h_{1M}(k)\equiv 2A_{1M}f_{\rm mod}(k),
  \label{eq:h1M_def}
\end{align}
where $A_{1M}$ are the dipolar modulation coefficients and $f_{\rm mod}(k)$ describes the scale dependence.
{
Alternatively, the dipolar modulation can be characterized by an overall amplitude $A$ and a preferred direction $\hat{d}$~\cite{Hoftuft:2009rq,Aiola:2015rqa}.
The coefficients $A_{1M}$ are then given by
}
\begin{align}
  A_{1M}
  =
  \frac{4\pi}{3}\,A\,Y_{1M}^{\ast}(\hat{d}).
  \label{eq:A1M_axisym}
\end{align}


This model is motivated by the hemispherical power asymmetry reported in the CMB temperature fluctuations.  Analyses of large angular scales have reported evidence at approximately the $3\sigma$ level for a nonzero dipolar modulation with amplitude $A\simeq0.06$--$0.07$, together with indications that the modulation amplitude decreases toward smaller angular scales~\cite{Gordon:2006ag,Hoftuft:2009rq,Aiola:2015rqa,Planck:2019evm}.


Ref.~\cite{Aiola:2015rqa} modeled the angular-scale dependence as a power law, $A(\ell)\propto(\ell/60)^{\beta}$, and obtained best-fit indices of approximately $\beta=-0.5$ to $-0.7$, depending on the maximum multipole included.  Motivated by this result, we adopt $f_{\rm mod}(k) = (k/k_A)^{-0.5}$ with $k_A = 0.005 \,\rm Mpc^{-1}$ as a representative scale-dependent model for the dipolar modulation.
The correspondence between $\ell$ and $k$ is only approximate, so this choice should be regarded as a CMB-motivated phenomenological parametrization rather than a direct translation of the CMB fit.

For the quadrupolar case, 
{
we express the modulation amplitude $h_{2M}$ by
}
\begin{align}
  h_{2M}(k)\equiv g_{2M}f_{\rm mod}(k),
  \label{eq:h2M_def}
\end{align}
where $g_{2M}$ characterizes the quadrupolar power asymmetry.

{A commonly used parametrization of axisymmetric quadrupolar power asymmetry introduces an amplitude $g_{\ast}$ and a preferred direction $\hat{d}$~\cite{Ackerman:2007nb,Sugiyama:2017ggb}.
In this parametrization, the matter power spectrum can be written as
}
\begin{align}
  P_{\rm m}^{(2)}(k,\hat{k})
  =
  \bar{P}_{\rm m}(k)
  \left[
  1+\frac{2}{3}g_{\ast}f_{\rm mod}(k)\,
  \mathcal{L}_2(\hat{k}\cdot\hat{d})
  \right].
  \label{eq:Pm2_axisym}
\end{align}
Using the spherical-harmonic addition theorem and comparing Eq.~\eqref{eq:Pm2_axisym} with Eqs.~\eqref{eq:Pmn_unified} and \eqref{eq:h2M_def}, we obtain
\begin{align}
  g_{2M}
  =
  \frac{8\pi}{15}\,g_{\ast}\,
  Y_{2M}^{\ast}(\hat{d}).
  \label{eq:g2M_axisym}
\end{align}
Observationally, quadrupolar power asymmetry is constrained by the CMB at the level of $|g_{\ast}|\lesssim \mathcal{O}(10^{-2})$, while large-scale-structure analyses currently give weaker bounds of order $\mathcal{O}(10^{-1})$~\cite{Kim:2013gka,Ramazanov:2016gjl,Planck:2019evm,Sugiyama:2017ggb}.
Forecasts for future surveys suggest that sensitivities comparable to those of the CMB may be achievable~\cite{Pullen:2007tu,Shiraishi:2016omb,Shiraishi:2016wec,Shiraishi:2020pea}.
The scale dependence of the quadrupolar amplitude is model dependent.  For example, anisotropic inflation produces a dependence through the number of e-folds $N(k)$, whereas other early-Universe scenarios can generate different $k$ dependences~\cite{Watanabe:2010fh,Soda:2012zm,Bartolo:2012sd,Agullo:2022klq}.
Since present observations do not select a unique scale dependence, we adopt the simple phenomenological form $f_{\rm mod}(k)=(k/k_g)^\alpha$ with $k_g = 0.05\,\rm Mpc^{-1}$ and examine several representative values of the power-law index $\alpha$.  This also facilitates comparison with CMB searches that have tested the discrete choices $\alpha=0,\pm1,\pm2$~\cite{Shiraishi:2016omb,Ramazanov:2016gjl,Shiraishi:2016wec,Shiraishi:2020pea}.

A key difference from the dipolar case is that the quadrupolar asymmetry also induces an anisotropic-bias contribution in galaxy clustering \cite{Shiraishi:2023zda}.
To make this explicit, we introduce the model-dependent probe kernels
\begin{align}
  K_{\rm g}^{(n)}(\bm{k},\hat{n})
  =
  K_{\rm g}(\mu)
  +\delta_{n2}\,\Delta K_{\rm g}(k,\hat{k}\cdot\hat{d})
  ,
  \label{eq:Kgn_def}
\end{align}
\begin{align}
  K_{\rm v}^{(n)}(k,\mu)=K_{\rm v}(k,\mu),
  \label{eq:Kvn_def}
\end{align}
where $\delta_{n2}$ is the Kronecker delta.
The correction $\Delta K_{\rm g}$ is present only for $n=2$ and is given by
\begin{align}
  \Delta K_{\rm g}(k,\hat{k}\cdot\hat{d})
  =
  \frac{1}{3}b_1^{(2)}g_{\ast}
  f_{\rm mod}(k)
  \mathcal{L}_2(\hat{k}\cdot\hat{d})
  ,
  \label{eq:DeltaKg_def}
\end{align}
with $b_1^{(2)}$ denoting the anisotropic-bias parameter.
The $N$-body halo simulations of Ref.~\cite{Masaki:2024hzn} find a negative anisotropic-bias coefficient whose magnitude increases with halo mass, 
{
together with an approximately linear relation between $b_1^{(2)}$ and $b_1-1$.  Motivated by this result, we adopt the following relation as the fiducial model in our forecasts:
}
\begin{align}
  b_1^{(2)}(z)=-0.5\,[b_1(z)-1].
  \label{eq:b12_model}
\end{align}
{
For the Euclid-like galaxy sample adopted in the Fisher analysis below, a corresponding calibration of $b_1^{(2)}$ is not available.  We therefore extrapolate the fitting relation calibrated for cluster-scale halos and use it as our fiducial model, since it provides a simulation-based estimate of the sign and approximate scaling of $b_1^{(2)}$ with $b_1$.  In the Fisher analysis, however, we allow the overall amplitude of this relation to vary as a nuisance parameter, while the velocity auto-spectrum is independent of $b_1^{(2)}$.  The forecast therefore does not require the normalization of the extrapolated relation to be known a priori.
}

Using the baseline kernels introduced in Sec.~\ref{sec:probes} and the model-dependent modulated matter power spectrum defined above, the observable power spectra can be written in the unified form:
\begin{align}
  P^{{\rm XY}(n)}(\bm{k},\hat{n})
  =
  K_{\rm X}^{(n)}(k,\mu)
  K_{\rm Y}^{(n)\ast}(k,\mu)
  P_{\rm m}^{(n)}(k,\hat{p}_n)
  ,
  \qquad
  {\rm X,Y}\in\{{\rm g,v}\}.
  \label{eq:PXYn_unified}
\end{align}

For the dipolar case ($n=1$), the correction due to the power asymmetry enters only through the matter power spectrum, and the observable kernels remain identical to the isotropic ones.
For the quadrupolar case ($n=2$), the correction enters both through the matter power spectrum and through the additional correction $\Delta K_{\rm g}$ to the galaxy kernel.
{
In evaluating Eq.~\eqref{eq:PXYn_unified}, we retain only terms linear in $g_{2M}$ or $g_*$ and neglect all terms of quadratic or higher order.
}
The velocity auto-spectrum receives no anisotropic-bias contribution
and therefore provides direct access to the primordial quadrupolar asymmetry.

\section{BipoSH Decomposition}
\label{sec:biposh}

To characterize the angular dependence of the observable power spectra, we employ the Bipolar Spherical Harmonics (BipoSH) formalism~\cite{Shiraishi:2016wec,Bartolo:2017sbu,Sugiyama:2017ggb,Akitsu:2019evv,Shiraishi:2020pea,Shiraishi:2023zda}.
This basis is natural when the spectra depend separately on the two directions $\hat{k}$ and $\hat{n}$, as is the case in the presence of power asymmetry.

\subsection{Definition}

The BipoSH basis functions are defined by~\cite{Varshalovich:1988ye}
\begin{align}
  X_{\ell\ell'}^{LM}(\hat{k},\hat{n})
  \equiv
  \left\{
  Y_{\ell}(\hat{k})\otimes Y_{\ell'}(\hat{n})
  \right\}_{LM}
  =
  \sum_{mm'}
  \mathcal{C}^{LM}_{\ell m \ell' m'}
  Y_{\ell m}(\hat{k})Y_{\ell' m'}(\hat{n}),
  \label{eq:BipoSH_def}
\end{align}
where $\mathcal{C}^{LM}_{\ell m \ell' m'}$ are the Clebsch--Gordan coefficients.

These basis functions satisfy the orthogonality relation
\begin{align}
  \int d^2\hat{k}\int d^2\hat{n}\,
  X_{\ell\ell'}^{LM}(\hat{k},\hat{n})
  X_{\tilde{\ell}\tilde{\ell}'}^{\tilde{L}\tilde{M}\ast}(\hat{k},\hat{n})
  =
  \delta_{\ell,\tilde{\ell}}
  \delta_{\ell',\tilde{\ell}'}
  \delta_{L,\tilde{L}}
  \delta_{M,\tilde{M}}.
  \label{eq:BipoSH_orthogonality}
\end{align}

The observable power spectra are decomposed as
\begin{align}
  P^{{\rm XY}(n)}(\bm{k},\hat{n})
  =
  \sum_{LM\ell\ell'}
  \,_{\rm XY}\pi_{\ell\ell'}^{LM\,(n)}(k)\,
  X_{\ell\ell'}^{LM}(\hat{k},\hat{n}).
  \label{eq:PXY_BipoSH}
\end{align}

In the statistically isotropic limit, only the $L=0$ BipoSH sector is non-vanishing, and the BipoSH expansion reduces to the ordinary Legendre expansion in Eq.~\eqref{eq:PXY_legendre}.
{
In the presence of power asymmetry, the $L=1$ and $L=2$ BipoSH sectors become non-vanishing for the dipolar and  quadrupolar cases, respectively.
}


\subsection{BipoSH coefficients for dipolar and quadrupolar power asymmetries}
\label{sec:biposh_unified}

Using the unified form of the modulated matter power spectrum in Eq.~\eqref{eq:Pmn_unified}, the BipoSH coefficients can be written in a unified manner for both $n=1$ and $n=2$ as
\begin{align}
  \,_{\rm XY}\pi_{\ell\ell'}^{LM\,(n)}(k)
  &=
  P_{\ell}^{\rm XY}(k)
  \frac{4\pi}{2\ell+1}
  H_{\ell\ell 0}^{-1}
  \delta_{\ell,\ell'}\delta_{L,0}\delta_{M,0}
  \nonumber\\
  &\quad
  +
  \mathcal{P}_{\ell_n}^{{\rm XY}(n)}(k)
  \sqrt{
  4\pi
  \left(
  \frac{2\ell+1}{2\ell'+1}
  \right)^{(-1)^{n+1}}
  }
  H_{\ell\ell' n}\,
  h_{nM}(k)\,
  \delta_{L,n},
  \label{eq:BipoSH_unified}
\end{align}
where
\begin{align}
  \ell_n \equiv
  \begin{cases}
  \ell, & n=1,\\
  \ell', & n=2,
  \end{cases}
  \label{eq:elln_def}
\end{align}

\begin{align}
  \mathcal{P}_{\ell_n}^{{\rm XY}(n)}(k)
  =
  P_{\ell_n}^{\rm XY}(k)
  +
  \delta_{n2}\,\Delta P_{\ell_n}^{\rm XY}(k).
  \label{eq:calP_unified}
\end{align}
and
\begin{align}
  H_{\ell_1\ell_2\ell_3}
  \equiv
  \begin{pmatrix}
  \ell_1 & \ell_2 & \ell_3\\
  0 & 0 & 0
  \end{pmatrix}.
  \label{eq:H_def}
\end{align}
The parentheses in Eq.~\eqref{eq:H_def} denote the Wigner $3j$ symbol.

{
The definition of $\ell_n$ in Eq.~\eqref{eq:elln_def} reflects whether the modulation depends on $\hat{n}$ ($n=1$) or $\hat{k}$ ($n=2$).  For $n=2$, $\mathcal{P}_{\ell'}^{{\rm XY}(2)}(k)$ additionally includes the anisotropic-bias correction $\Delta P_{\ell'}^{\rm XY}(k)$.
}

The correction terms $\Delta P_{\ell}^{\rm XY}(k)$ are nonzero only for the quadrupolar case {($n=2$)}.
{
For the galaxy auto-spectrum, the corrections to the monopole and
quadrupole are nonzero, while the correction to the hexadecapole
vanishes:
}
\begin{align}
  \Delta P_{0}^{\rm gg}(k)
  &=
  \left(
  b_1 b_1^{(2)}+\frac{1}{3}b_1^{(2)}f
  \right)\bar{P}_{\rm m}(k),
  \\
  \Delta P_{2}^{\rm gg}(k)
  &=
  \frac{2}{3}b_1^{(2)}f\,\bar{P}_{\rm m}(k),
  \\
  \Delta P_{4}^{\rm gg}(k)
  &=
  0.
  \label{eq:DeltaP_gg}
\end{align}
{
For the density--velocity cross-spectrum, the correction is nonzero
only for the dipole:
}
\begin{align}
  \Delta P_{1}^{\rm gv}(k)
  &=
  -\,i\,\frac{aHf}{2k}\,b_1^{(2)}\,\bar{P}_{\rm m}(k),
  \\
  \Delta P_{3}^{\rm gv}(k)
  &=
  0.
  \label{eq:DeltaP_gv}
\end{align}
{
For the velocity auto-spectrum, the corrections to both the monopole
and quadrupole vanish:
}
\begin{align}
  \Delta P_{0}^{\rm vv}(k)
  &=
  0,
  \\
  \Delta P_{2}^{\rm vv}(k)
  &=
  0.
  \label{eq:DeltaP_vv}
\end{align}

{
Thus, Eq.~\eqref{eq:BipoSH_unified} describes the BipoSH coefficients for both the dipolar and quadrupolar cases.  The two cases are distinguished by the value of $n$, which determines the directional argument $\hat{p}_n$ through Eq.~\eqref{eq:pn_def}, and by the anisotropic-bias correction $\Delta P_{\ell}^{\rm XY}(k)$, which is present only for $n=2$.
}

\section{Fisher Forecast}
\label{sec:fisher}

We use the Fisher matrix formalism to quantify the detectability of the dipolar and quadrupolar power asymmetries.
Following previous studies based on the BipoSH expansion~\cite{Tegmark:1996bz,Shiraishi:2016wec,Shiraishi:2020pea,Shiraishi:2023zda,Minato:2025ozy},
we present the formulation of the Fisher analysis below.

{
For the dipolar asymmetry, we treat $A_{1M}$ as the only free parameter
in the Fisher analysis and estimate its expected uncertainty from the
$L=1$ BipoSH coefficients.  The corresponding parameter set is
}
\begin{align}
\theta_i \in \{A_{1M}\}.
\end{align}

{
For the quadrupolar asymmetry, we treat $g_{2M}$ and $b_1^{(2)}$ as
free parameters in the Fisher analysis and estimate their joint
constraints from the $L=2$ BipoSH coefficients.  The corresponding
parameter set is
}
\begin{align}
\theta_i \in \{ g_{2M},\, b_1^{(2)} \},
\end{align}
where $g_{2M}$ represents the primordial quadrupolar power asymmetry and $b_1^{(2)}$ denotes the anisotropic bias parameter.
Since $b_1$ is fixed by the isotropic sector, marginalizing over $\alpha_b$ is equivalent to marginalizing over $b_1^{(2)}$.
Here, we write $b_1^{(2)}(z)=\alpha_b[b_1(z)-1]$, assume that $\alpha_b=b_1^{(2)}/(b_1-1)$ is redshift independent, and set its fiducial value to $\alpha_b^{\rm fid}=-0.5$.  Thus, the numerical analysis marginalizes over $\alpha_b$, keeping the assumed redshift dependence of $b_1^{(2)}(z)$ fixed while allowing its overall amplitude to vary.


{
In addition to the Fisher parameters introduced above, the power spectra depend on the isotropic linear bias $b_1$
and growth rate $f$. Ref.~\cite{Minato:2025ozy} showed that, for small modulation amplitudes, marginalizing over such isotropic parameters changes the constraints on the modulation amplitude only at second order. We therefore fix $b_1$ and $f$ in the present Fisher analysis.
}

The Fisher matrix is defined as
\begin{align}
F_{ij}^{(n)}
&= \sum_{k_1,k_2}^{k_{\rm max}}\sum_{\mathcal{C}_1,\mathcal{C}_2}
\frac{\partial \,_{\rm X_1Y_1}\pi_{\ell_1\ell_1^\prime}^{L_1M_1\,(n)}(k_1)^\ast}{\partial \theta_i^\ast}
\nonumber\\
&\quad\times
\left\langle
\,_{\rm X_1Y_1}\pi_{\ell_1\ell_1^\prime}^{L_1M_1\,(n)}(k_1)^\ast
\,_{\rm X_2Y_2}\pi_{\ell_2\ell_2^\prime}^{L_2M_2\,(n)}(k_2)
\right\rangle_c^{-1}
\frac{\partial \,_{\rm X_2Y_2}\pi_{\ell_2\ell_2^\prime}^{L_2M_2\,(n)}(k_2)}{\partial \theta_j}.
\end{align}
{The quantity with bracket is the covariance matrix for BipoSH coefficients (see below).} 
Here, $\mathcal{C}_i$ collectively denotes the set of indices $(\ell_i,\ell_i^\prime,L_i,M_i,{\rm X}_i,{\rm Y}_i)$.

In the presence of observational noise, the observed power spectra are given by
\begin{align}
\widetilde{P}^{\rm XY}(k) = P^{\rm XY}(k) + N^{\rm XY}(k).
\end{align}
{The second term on the right-hand side represents the noise contribution.} 
For galaxy clustering, the noise term {$N^{\rm gg}$} is given by the shot noise~\cite{Feldman:1993ky}
\begin{align}
N^{\rm gg}(k) = \frac{1}{\bar{n}_{\rm g}},
\end{align}
where $\bar{n}_{\rm g}$ is the mean number density of galaxies.

For the cross and velocity spectra, we assume\footnote{The noise power $N^{\rm vv}$ in Eq.~(5.6) is obtained by converting the noise power of the kSZ temperature fluctuation into that of the line-of-sight peculiar velocity using the relation given in footnote~\ref{fn:ksz_velocity_relation}.}
\begin{align}
N^{\rm gv}(k) = 0, \qquad
N^{\rm vv}(k) = \left(1 + R_N^2\right) \frac{(faH\sigma_d)^2}{\bar{n}_{\rm v}},
\end{align}
Here, $R_N$ is the ratio of the reconstruction noise per velocity tracer to the cosmological rms line-of-sight velocity, and $\bar n_{\rm v}$ is the mean number density of the velocity tracers.
The quantity $\sigma_d$ is the one-dimensional rms linear displacement, defined from the fiducial isotropic matter power spectrum by
\begin{align}
  \sigma_d^2
  =
  \frac{1}{6\pi^2}\int dq\,\bar{P}_{\rm m}(q),
  \qquad
  \sigma_v=aHf\sigma_d,
\end{align}
so that $\sigma_v^2=\langle v_\parallel^2\rangle$ in linear theory.
{
Here, we assume that the effective velocity-noise power is independent of wavenumber.
}
Realistic reconstruction noise can depend on scale, redshift, tracer selection, optical-depth calibration, and higher-order correlations~\cite{Deutsch:2017ybc,Smith:2018bpn,Munchmeyer:2018eey,Giri:2020pkk,Contreras:2022zdz,McCarthy:2024nik}. 
For the numerical forecasts, we approximate $\sigma_v$ as redshift independent and set $\sigma_v=300\,{\rm km\,s^{-1}}$, following the linear-theory estimate of Ref.~\cite{Okumura:2021xgc} based on Ref.~\cite{Vlah:2012ni}.


The covariance matrix of the power spectra is given by
\begin{align}
&\left\langle P^{\rm X_1Y_1}(\bm k_1, \hat{n}_1) \,P^{\rm X_2Y_2}(\bm k_2, \hat{n}_2) \right\rangle_c \nonumber\\
&= 4\pi \frac{\delta_{k_1, k_2}}{N_{k_1}} \sum_{J_1, J_2} \mathcal{L}_{J_1}(\hat{k}_1\cdot\hat{n}_1) \mathcal{L}_{J_2}(\hat{k}_2\cdot\hat{n}_2)
\times 4\pi \delta_{\rm D}^{(2)}(\hat{n}_1-\hat{n}_2) \nonumber\\
&\quad \times \left[ \widetilde{P}^{\rm X_1X_2}_{J_1}(k_1) \widetilde{P}^{\rm Y_1Y_2}_{J_2}(k_1) \delta_{\rm D}^{(2)}(\hat{k}_1+\hat{k}_2) + \widetilde{P}^{\rm X_1Y_2}_{J_1}(k_1) \widetilde{P}^{\rm X_2Y_1}_{J_2}(k_1) \delta_{\rm D}^{(2)}(\hat{k}_1-\hat{k}_2)\right],
\end{align}
where $N_k$ denotes the number of independent Fourier modes in the shell at wavenumber $k$.

Using the orthogonality of the BipoSH basis {and assuming Gaussian statistics at large scales}, this covariance can be projected into the covariance of the BipoSH coefficients as~\cite{Shiraishi:2020pea}
\begin{align}
\left\langle{}_{\rm X_1Y_1}\pi_{\ell_1\ell_1^\prime}^{L_1M_1}(k_1)^\ast \,_{\rm X_2Y_2}\pi_{\ell_2\ell_2^\prime}^{L_2M_2}(k_2)\right\rangle_c
=
\delta_{L_1,L_2}\delta_{M_1,M_2}\frac{\delta_{k_1, k_2}}{N_{k_1}}
\Theta_{\ell_1\ell_1^\prime \ell_2\ell_2^\prime}^{L_1,\,\rm X_1Y_1X_2Y_2}(k_1).
\end{align}
The explicit form of $\Theta$ is given by~\cite{Minato:2025ozy}
\begin{align}
  &\Theta_{\ell_1\ell_1^\prime \ell_2\ell_2^\prime}^{L,\,\rm X_1Y_1X_2Y_2}(k)
  \nonumber\\
  =&~ 16\pi^2 (-1)^{\ell_1+\ell_2^\prime}
  \sum_{\ell,J,J^\prime}
  \left[
  \widetilde{P}_{J}^{\rm X_1X_2}(k)
  \widetilde{P}_{J^\prime}^{\rm Y_1Y_2}(k)
  +(-1)^{\ell_2}
  \widetilde{P}_{J}^{\rm X_1Y_2}(k)
  \widetilde{P}_{J^\prime}^{\rm X_2Y_1}(k)
  \right]
  \nonumber\\
  &\times
  (2\ell+1)
  \sqrt{
  (2\ell_1+1)(2\ell_1^\prime+1)
  (2\ell_2+1)(2\ell_2^\prime+1)
  }
  H_{JJ^\prime \ell}^2
  H_{\ell_1\ell_2\ell}
  H_{\ell_1^\prime \ell_2^\prime \ell}
  \left\{
  \begin{matrix}
  L&\ell_2&\ell_2^\prime\\
  \ell&\ell_1^\prime&\ell_1
  \end{matrix}
  \right\}.
  \label{eq:Theta_explicit}
\end{align}
The braces in Eq.~\eqref{eq:Theta_explicit} denote the Wigner $6j$ symbol.

Finally, the Fisher matrix can be written as
\begin{align}
F_{ij}^{(n)}
= V_{\rm eff}\int_{k_{\rm min}}^{k_{\rm max}}\frac{k^2dk}{2\pi^2}
\sum_{\substack{
\ell_1, \ell_1^\prime, \ell_2, \ell_2^\prime, L, M, \\
\rm X_1, Y_1, X_2, Y_2
}}
\frac{\partial \,_{\rm X_1Y_1}\pi_{\ell_1\ell_1^\prime}^{LM\,(n)}(k)^\ast}{\partial \theta_i^\ast}
\left(\Theta^{-1}\right)_{\ell_1\ell_1^\prime \ell_2\ell_2^\prime}^{L,\rm X_1Y_1X_2Y_2}(k)
\frac{\partial \,_{\rm X_2Y_2}\pi_{\ell_2\ell_2^\prime}^{LM\,(n)}(k)}{\partial \theta_j}.
\end{align}
{
Here, $V_{\rm eff}$ is the effective survey volume accounting for the Euclid--CMB overlap. See Sec.~\ref{subsec:survey_setup} for details.
}

{In our Fisher matrix analysis in Sec.~\ref{sec:results}, we use multiple redshift slices of the spectroscopic survey to constrain the power asymmetry. Assuming that the data sets in different redshift slices are statistically independent, the total Fisher matrix is given by} 
\begin{align}
F_{ij}^{\mathrm{tot},(n)} = \sum_a F_{ij}^{(n)}(z_a).
\end{align}
{
The unmarginalized and marginalized $1\sigma$ errors on $\theta_i$ are given by
}
\begin{align}
&\Delta\theta_i^{\rm unmarg}
=
\frac{1}{\sqrt{F_{ii}^{\rm tot}}},
\\
&\Delta\theta_i^{\rm marg}
=
\sqrt{\left[(F^{\rm tot})^{-1}\right]_{ii}},
\end{align}
{
respectively}
{, where the unmarginalized error fixes all other parameters to their fiducial values, while the marginalized error accounts for their uncertainties.}

\section{Results}
\label{sec:results}

\subsection{Survey and noise setup}
\label{subsec:survey_setup}

{For our fiducial survey setup, we adopt a Euclid-like spectroscopic galaxy sample divided into four redshift bins, following the survey specifications of Ref.~\cite{Euclid:2019clj}. Euclid's wide sky coverage and large spectroscopic volume make it well suited for the joint analysis of galaxy clustering and peculiar velocities considered here.}

The galaxy bias, number density, and comoving volume used in the forecasts are summarized in Table~\ref{tab:euclid_survey}.  The galaxy number density enters the shot-noise contribution, and the volume determines both the normalization of the Fisher integral and the fundamental wavenumber of each bin.  We use the Planck 2018 cosmological parameters for the fiducial background and linear matter power spectrum~\cite{Planck:2018vyg}.

\begin{table}[t]
\centering
\begin{tabular}{c c c c}
\hline
$z$ & $b_1$ & $\bar n_{\rm g}\,[h^3{\rm Mpc}^{-3}]$ & $V\,[h^{-3}{\rm Gpc}^3]$ \\
\hline
1.00 & 1.46 & $6.86\times 10^{-4}$ & 7.94 \\
1.20 & 1.61 & $5.58\times 10^{-4}$ & 9.15 \\
1.40 & 1.75 & $4.21\times 10^{-4}$ & 10.05 \\
1.65 & 1.90 & $2.61\times 10^{-4}$ & 16.22 \\
\hline
\end{tabular}
\caption{Euclid survey specifications adopted in the Fisher forecasts.  The entries are the central redshift of each bin, the linear galaxy bias, the mean galaxy number density, and the comoving volume of the bin~\cite{Euclid:2019clj}.}
\label{tab:euclid_survey}
\end{table}

For the velocity field, we assume a kSZ velocity-reconstruction measurement using CMB maps with Simons Observatory-like sky coverage~\cite{SimonsObservatory:2018koc}.  Using the scale-independent noise model introduced in Sec.~5, we adopt $R_N=10$ and $\bar n_{\rm v}=\bar n_{\rm g}$ for the fiducial {setup.}

{Note that} the reconstructed velocity field is available only in the angular overlap between the galaxy survey and the CMB map.  We denote by $\eta_{\rm cov}$ the fraction of the usable Euclid footprint that overlaps the CMB map and use $\eta_{\rm cov}=0.7$ as a fiducial benchmark, motivated by the substantial Euclid--CMB sky overlap illustrated in Fig.~2 of Ref.~\cite{Euclid:2021cmbjoint}.  This value is intended as a simple overlap assumption, not as the result of a detailed survey-mask calculation. 
{
Following the treatment of partial survey overlap in Ref.~\cite{Okumura:2021xgc}, we assign an effective volume to each contribution according to the sky area over which the corresponding fields are jointly available.  
}
In the forecasts, the fundamental wavenumber is set by the effective volume of the relevant region, $V_{\rm eff}$, as $k_{\rm min}=2\pi/V_{\rm eff}^{1/3}$.  Galaxy-only information uses the full Euclid volume, $V_{\rm eff}=V$, 
while information from the density--velocity cross-spectrum and velocity auto-spectrum uses the overlap volume, $V_{\rm eff}=\eta_{\rm cov}V$.  For the joint galaxy--velocity analyses, we add the galaxy-only Fisher information from the non-overlap region to the joint Fisher information from the overlap region.

{Below, we} present the Fisher forecasts obtained from the galaxy-density and line-of-sight velocity fields.  We first present the dipolar forecast in Sec.~\ref{subsec:results_dipole} and then turn to the quadrupolar model in Sec.~\ref{subsec:results_quadrupole}, for which the anisotropic-bias response introduces an additional parameter.  For the {latter case,}
we subsequently examine the dependence on the maximum wavenumber, the velocity-reconstruction noise, and the CMB--Euclid overlap fraction.  Unless otherwise stated, the results use the survey and noise assumptions described above 
{
and adopt $k_{\rm max}=0.1\,h\,{\rm Mpc}^{-1}$.
}

\subsection{Dipolar power asymmetry}
\label{subsec:results_dipole}

We first consider the dipolar model, whose only power-asymmetry parameter is $A_{1M}$ in our phenomenological parametrization.
We use $f_{\rm mod}(k)=1$ and $f_{\rm mod}(k)=(k/k_A)^{-0.5}$ with $k_A=0.005\,{\rm Mpc}^{-1}$.  The latter represents a CMB-motivated modulation that is stronger on large scales.

\begin{figure}[t]
  \centering
  \begin{minipage}[t]{0.49\textwidth}
  \centering
  \includegraphics[width=\linewidth]{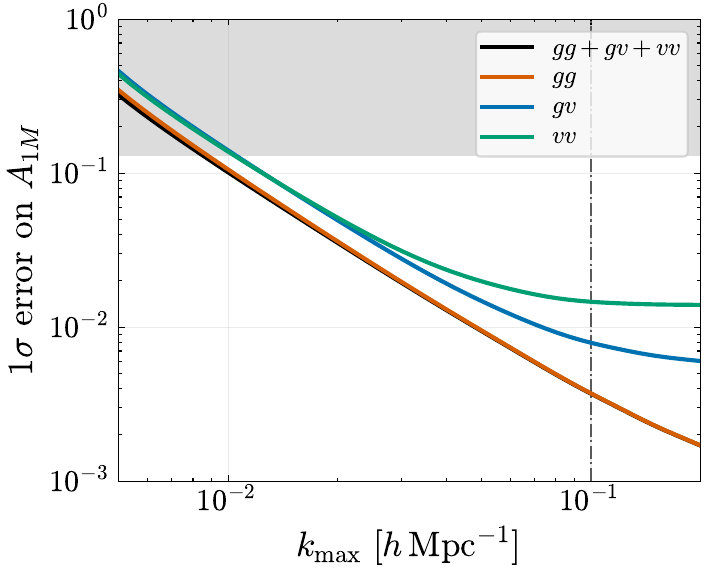}
  \subcaption{$f_{\rm mod}(k)=1$.}
  \end{minipage}
  \hfill
  \begin{minipage}[t]{0.49\textwidth}
  \centering
  \includegraphics[width=\linewidth]{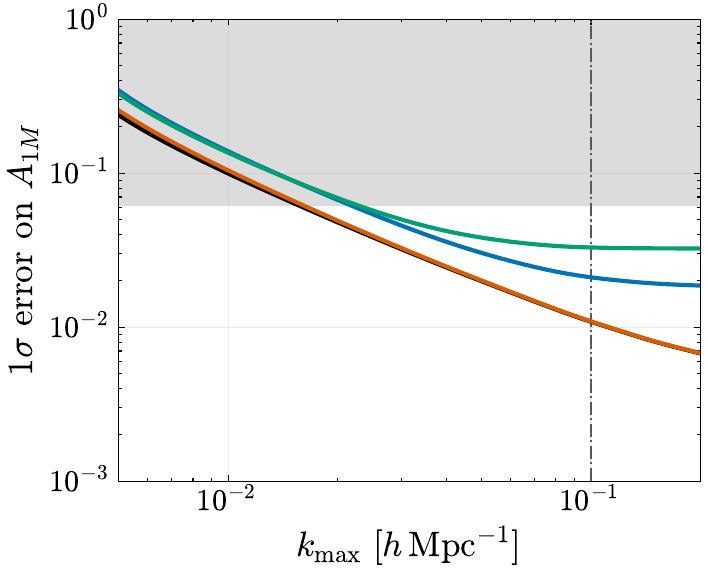}
  \subcaption{$f_{\rm mod}(k)=(k/k_A)^{-0.5}$.}
  \end{minipage}
  \caption{Expected $1\sigma$ errors on the dipolar modulation coefficient $A_{1M}$.  The black, orange, blue, and green curves show the joint $gg+gv+vv$ result and the individual $gg$, $gv$, and $vv$ results, respectively.  The gray shaded regions indicate errors larger than the representative CMB-inferred amplitudes adopted here.  After translating the scalar modulation amplitudes reported in Ref.~\cite{Aiola:2015rqa} to the preferred-direction frame using $A_{10}=\sqrt{4\pi/3}\,A$, the values are $A_{10}\simeq0.13$ in panel (a) and $A_{10}\simeq0.061$ in panel (b).  The dot-dashed vertical lines mark $k_{\rm max}=0.1\,h\,{\rm Mpc}^{-1}$.}
  \label{fig:dipole_constraints}
\end{figure}

Figure~\ref{fig:dipole_constraints} shows the forecast $1\sigma$ error on $A_{1M}$ as a function of $k_{\rm max}$.  The scale-independent model gives tighter constraints because the modulation remains unsuppressed up to $k_{\rm max}$.  At $k_{\rm max}=0.1\,h\,{\rm Mpc}^{-1}$, we find
\begin{align}
  \Delta A_{1M}^{\rm joint} &\simeq 3.7\times10^{-3}, &
  \Delta A_{1M}^{\rm gg} &\simeq 3.7\times10^{-3}, \\
  \Delta A_{1M}^{\rm gv} &\simeq 7.9\times10^{-3}, &
  \Delta A_{1M}^{\rm vv} &\simeq 1.5\times10^{-2}.
  \label{eq:dipole_errors_const}
\end{align}
Thus, at this representative scale, the $\mathrm{gv}$ and $\mathrm{vv}$ errors are about $2.1$ and $4.1$ times larger than the density-only error, respectively, while the full joint result is essentially identical to the $\mathrm{gg}$ result.  
The close agreement between the joint and $\mathrm{gg}$ curves over the displayed $k_{\rm max}$ range shows that this conclusion is not specific to the representative value $k_{\rm max}=0.1\,h\,{\rm Mpc}^{-1}$.
For $f_{\rm mod}(k)=(k/k_A)^{-0.5}$, the corresponding errors are
\begin{align}
  \Delta A_{1M}^{\rm joint} &\simeq 1.1\times10^{-2}, &
  \Delta A_{1M}^{\rm gg} &\simeq 1.1\times10^{-2}, \\
  \Delta A_{1M}^{\rm gv} &\simeq 2.1\times10^{-2}, &
  \Delta A_{1M}^{\rm vv} &\simeq 3.3\times10^{-2}.
  \label{eq:dipole_errors_scale}
\end{align}
For the scale-dependent dipole model, the corresponding error ratios are
\begin{align}
  \frac{\Delta A_{1M}^{\rm gv}}{\Delta A_{1M}^{\rm gg}}
  \simeq1.9,
  \qquad
  \frac{\Delta A_{1M}^{\rm vv}}{\Delta A_{1M}^{\rm gg}}
  \simeq3.0.
\end{align}
Combining all three spectra again gives only a negligible improvement.  The negative tilt suppresses the relative contribution of the increasingly numerous high-$k$ modes, so the constraint improves more slowly with $k_{\rm max}$ and remains weaker than in the scale-independent case over the range shown.
For both scale dependences, $\mathrm{gg}$ dominates and the joint constraint is close to the density-only result.
The velocity spectra therefore mainly provide a cross-check with different systematics, since a coherent signal in $\mathrm{gg}$, $\mathrm{gv}$, and $\mathrm{vv}$ would be harder to attribute to a probe-specific systematic.


\subsection{Quadrupolar power asymmetry}
\label{subsec:results_quadrupole}

\begin{figure}[t]
  \centering
  \begin{minipage}[t]{0.49\textwidth}
  \centering
  \includegraphics[width=\linewidth]{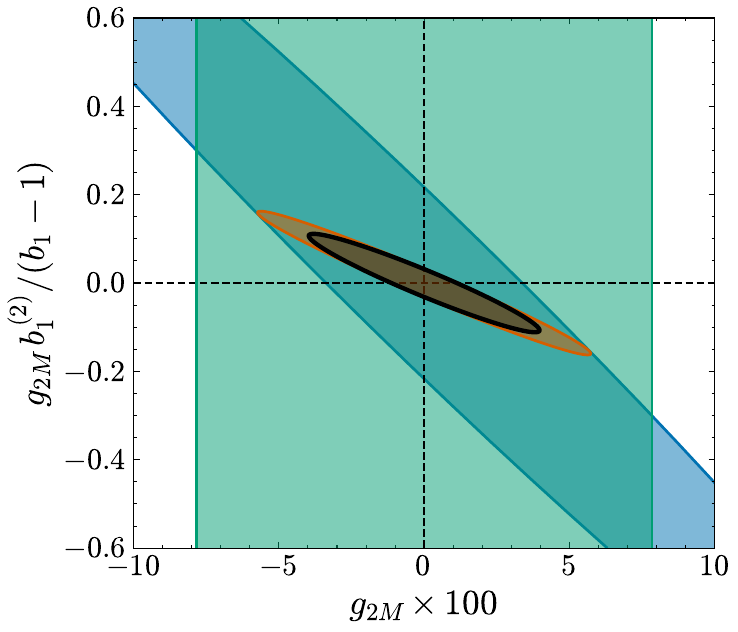}
  \subcaption{$f_{\rm mod}(k)=1$.}
  \end{minipage}
  \hfill
  \begin{minipage}[t]{0.49\textwidth}
  \centering
  \includegraphics[width=\linewidth]{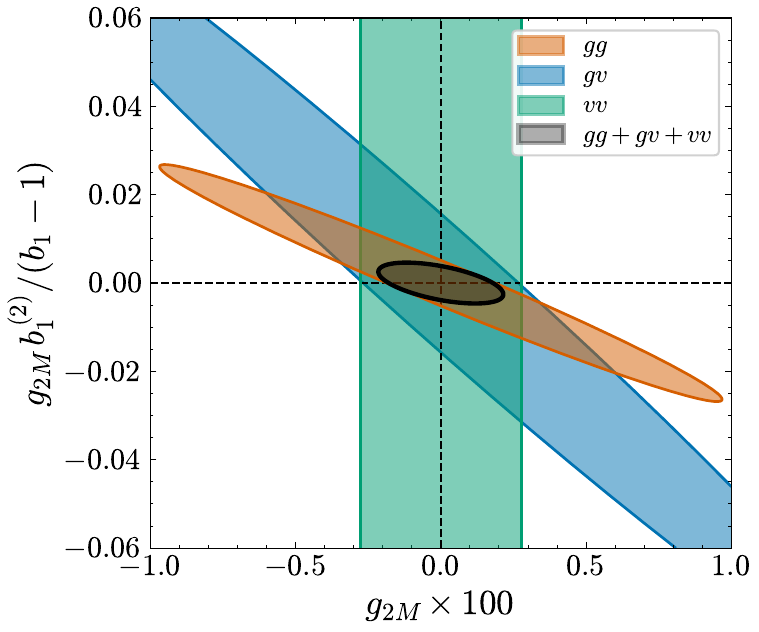}
  \subcaption{$f_{\rm mod}(k)=(k/k_g)^{-2}$.}
  \end{minipage}
  \caption{Forecast constraints in the $(g_{2M},\,g_{2M}b_1^{(2)}/(b_1-1))$ plane.  The red, blue, green, and black regions show the constraints from $\mathrm{gg}$, $\mathrm{gv}$, $\mathrm{vv}$, and the full $\mathrm{gg}+\mathrm{gv}+\mathrm{vv}$ analysis, respectively.  All contours correspond to $\Delta\chi^2=2.30$.  The velocity auto-spectrum produces an approximately vertical band because it is independent of $b_1^{(2)}$.}
  \label{fig:quadrupole_joint}
\end{figure}

{Since the quadrupolar asymmetry induces anisotropic-halo bias in the galaxy samples, we consider the simultaneous constraints on both the asymmetry parameter and bias parameter.} 
Figure~\ref{fig:quadrupole_joint} shows the joint constraints 
{on $g_{2M}$ and $b_1^{(2)}$} 
for the scale-independent model, $f_{\rm mod}(k)=1$, and for the 
{
scale-dependent model, $f_{\rm mod}(k)=(k/k_g)^{-2}$, with the pivot scale $k_g$ set to $0.05 \,{\rm Mpc^{-1}}$. 
}
{The vertical axis represents the anisotropic-bias contribution normalized by $b_1-1$, i.e., $g_{2M}b_1^{(2)}/(b_1-1)$, which is independent of redshift in the anisotropic-bias model (see Eq.~(\ref{eq:b12_model})).}

{The error contour}
from the galaxy auto-spectrum is {tilted and significantly stretched,}
reflecting the degeneracy between the primordial quadrupolar modulation and the anisotropic response of biased tracers.  
The density--velocity cross-spectrum alone also leaves a pronounced degeneracy, but its degeneracy direction in the $g_{2M}$--$b_1^{(2)}$ plane differs from that of the galaxy auto-spectrum.
The velocity auto-spectrum constrains $g_{2M}$ independently of $b_1^{(2)}$ and consequently appears as an approximately vertical band in Fig.~\ref{fig:quadrupole_joint}.

\begin{figure}[!htbp]
  \centering
  \begin{minipage}[t]{0.49\textwidth}
  \centering
  \includegraphics[width=\linewidth]{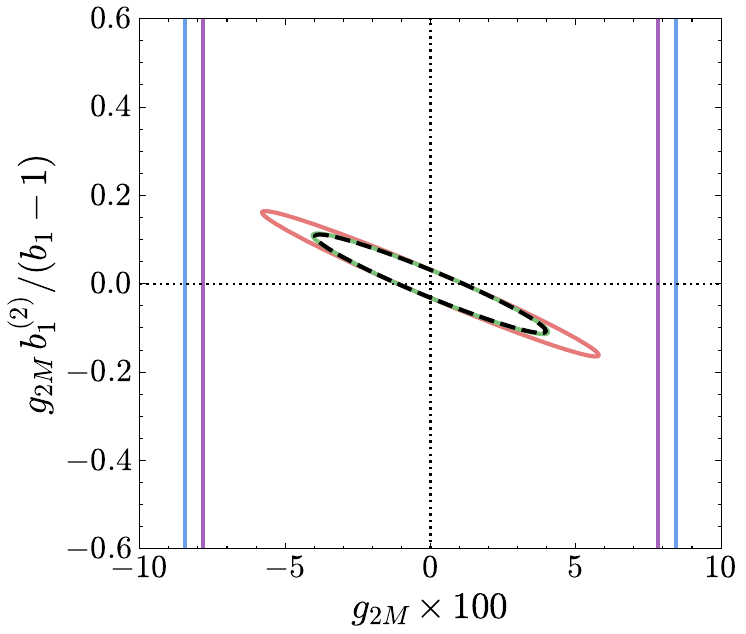}
  \subcaption{$f_{\rm mod}(k)=1$.}
  \end{minipage}
  \hfill
  \begin{minipage}[t]{0.49\textwidth}
  \centering
  \includegraphics[width=\linewidth]{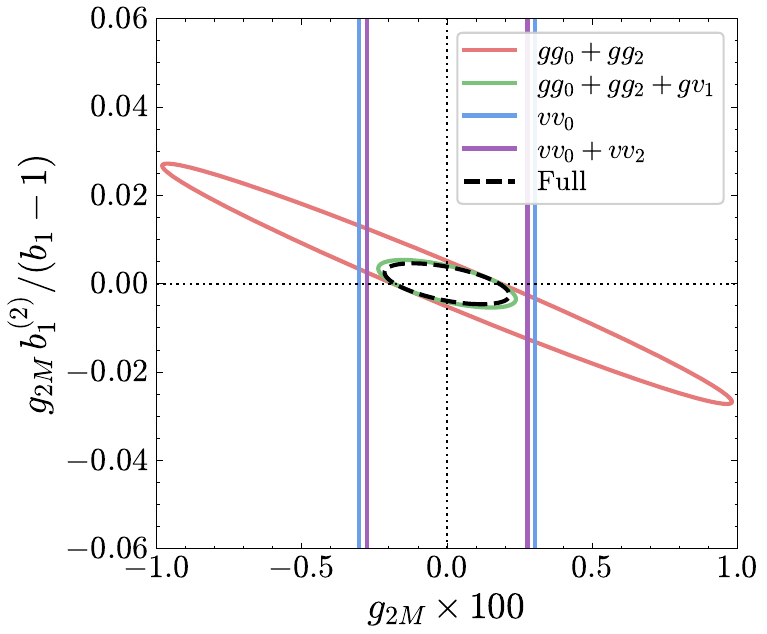}
  \subcaption{$f_{\rm mod}(k)=(k/k_g)^{-2}$.}
  \end{minipage}
  \caption{
Forecast constraints in the
$(g_{2M},\,g_{2M}b_1^{(2)}/(b_1-1))$ plane obtained from different
subsets of the quadrupolar BipoSH coefficients.
Panels (a) and (b) correspond to $f_{\rm mod}(k)=1$ and
$f_{\rm mod}(k)=(k/k_g)^{-2}$, respectively, with
$k_g=0.05\,{\rm Mpc}^{-1}$ in panel (b).
{In the legend, $\mathrm{XY}_{\ell'}$ represents the contribution from all nonvanishing BipoSH coefficients for a fixed $\ell'$, $\,_{\rm XY}\pi_{\ell\ell'}^{2M}$.}
The red and green contours show the constraints from
$\mathrm{gg}_0+\mathrm{gg}_2$ and
$\mathrm{gg}_0+\mathrm{gg}_2+\mathrm{gv}_1$, respectively.
The blue and purple vertical bands show the velocity-only constraints
from $\mathrm{vv}_0$ and $\mathrm{vv}_0+\mathrm{vv}_2$, respectively,
while the black dashed contours show the constraints obtained from 
{
the ``Full'' combination, which includes all $L=2$ BipoSH
coefficients that are non-vanishing within the linear, local
plane-parallel model adopted here.
}
All contour and band boundaries correspond to
$\Delta\chi^2=2.30$.
The velocity-only constraints are vertical because
$P^{\rm vv}$ is independent of $b_1^{(2)}$.
}
  \label{fig:multipole_breakdown}
\end{figure}

{To identify which BipoSH coefficients carry most of the constraining power in the joint analysis, Fig.~\ref{fig:multipole_breakdown} compares constraints obtained from different subsets of coefficients. In the figure legends, $\mathrm{XY}_{\ell'}$ denotes the set of quadrupolar BipoSH coefficients $\,_\mathrm{XY}\pi_{\ell\ell'}^{2M}$ with fixed $\ell'$ and all allowed $\ell$ satisfying $H_{\ell\ell'2}\neq0$. For example, $\mathrm{gg}_0+\mathrm{gg}_2+\mathrm{gv}_1$ denotes the combination of $\,_\mathrm{gg}\pi_{\ell0}^{2M}$, $\,_\mathrm{gg}\pi_{\ell2}^{2M}$, and $\,_\mathrm{gv}\pi_{\ell1}^{2M}$. The label ``Full'' includes all BipoSH coefficients that are nonzero at linear order: $\ell'=0,2,4$ for $\mathrm{gg}$, $\ell'=1,3$ for $\mathrm{gv}$, and $\ell'=0,2$ for $\mathrm{vv}$.}

{For both scale dependences shown in Fig.~\ref{fig:multipole_breakdown}, the combination of $\mathrm{gg}_0+\mathrm{gg}_2+\mathrm{gv}_1$ recovers most of the constraining power of the full set of coefficients. Under the fiducial survey and noise assumptions, its confidence contours are already close to those obtained using all coefficients. Adding the $\mathrm{vv}$ coefficients leads to only a modest further improvement. Nevertheless, the velocity auto-spectrum has a distinct advantage: it constrains $g_{2M}$ independently of the anisotropic-bias parameter $b_1^{(2)}$.}

Table~\ref{tab:scale_dependence_quadrupole} supplements the two cases displayed in Fig.~\ref{fig:quadrupole_joint} with the additional choices $\alpha=+1$ and $-1$.  Changing the tilt from $\alpha=+1$ to $-1$ improves the marginalized joint constraint from $\Delta g_{2M}=3.0\times10^{-2}$ to $1.0\times10^{-2}$, a factor of $3.0$.  Over the same change, 
the joint error {with $b_1^{(2)}$ fixed} improves only from $7.3\times10^{-3}$ to $5.5\times10^{-3}$, 
{corresponding to an improvement by a factor of about 1.3.}
{The larger improvement in the joint errors after marginalization indicates that a negative tilt enhances the signal at low-$k$ modes, where the velocity field is most sensitive, thereby increasing its constraining power and helping break the parameter degeneracies. Specifically, the density–velocity cross spectrum and the velocity auto-spectrum become more effective at breaking the degeneracy between anisotropic-bias and power-asymmetry parameters.} 
Correspondingly, 
{the factor by which marginalizing over $b_1^{(2)}$ increases the joint error}
decreases from a factor of about $4.1$ at $\alpha=+1$ to $1.8$ at $\alpha=-1$, and to about $1.2$ at $\alpha=-2$.

\begin{table}[t]
  \centering
  \caption{
  Representative quadrupole constraints at
  $k_{\rm max}=0.1\,h\,{\rm Mpc}^{-1}$ for different scale
  dependences $f_{\rm mod}(k)=(k/k_g)^\alpha$, with
  $R_N=10$ and $\eta_{\rm cov}=0.7$.
  All entries are $100\,\Delta g_{2M}$.
  The $\mathrm{vv}$ result is shown only once because it is independent
  of $b_1^{(2)}$ and is therefore unchanged by marginalization.
  {
  The penultimate column lists the approximate BOSS DR12 constraints
  used for comparison~\cite{Sugiyama:2017ggb}.
  }
  The last column gives the ideal cosmic-variance-limited CMB
  temperature and $E$-mode polarization forecast of
  Ref.~\cite{Shiraishi:2016omb}.
  }
  \label{tab:scale_dependence_quadrupole}

  \begin{talltblr}[
    label = none,
    entry = none,
    note{\textdagger} = {
      {
      The BOSS entries are obtained by rescaling the $1\sigma$ errors
  reported in Ref.~\cite{Sugiyama:2017ggb} to account for the
  anisotropic-bias response, with $b_1^{(2)}(z)$ fixed separately in each redshift bin according to
  Eq.~\eqref{eq:b12_model}.  They do not include marginalization over
  $b_1^{(2)}$.
      }
    },
  ]{
    colspec = {|c||c|ccc|ccc|c|c|},
    hline{1,3,Z},
    hline{2} = {3-8}{},
    colsep = 3.0pt,
    abovesep = 0pt,
    belowsep = 0pt,
  }
  \SetCell[r=2]{c} $\alpha$
  & \SetCell[r=2]{c} $\mathrm{vv}$
  & \SetCell[c=3]{c} $b_1^{(2)}$ unmarginalized
  & &
  & \SetCell[c=3]{c} $b_1^{(2)}$ marginalized
  & &
  & \SetCell[r=2]{c}
      {\shortstack{BOSS\TblrNote{\textdagger}~~\cite{Sugiyama:2017ggb}}}
  & \SetCell[r=2]{c}
      {\shortstack{CMB~\cite{Shiraishi:2016omb}
      }}
  \\
  &
  & $\mathrm{gg}$
  & $\mathrm{gv}$
  & $\mathrm{joint}$
  & $\mathrm{gg}$
  & $\mathrm{gv}$
  & $\mathrm{joint}$
  & &
  \\
  $+1$
  & 8.9
  & 0.73 & 2.7  & 0.73
  & 3.7  & 14   & 3.0
  & 0.90 & 0.083
  \\
  $0$
  & 5.2
  & 0.74 & 2.2  & 0.74
  & 3.8  & 12   & 2.6
  & 1.5  & 0.19
  \\
  $-1$
  & 1.5
  & 0.56 & 1.1  & 0.55
  & 2.8  & 5.9  & 1.0
  & 1.8  & 0.21
  \\
  $-2$
  & 0.18
  & 0.12 & 0.18 & 0.12
  & 0.64 & 0.99 & 0.14
  & 1.0  & 0.0037
  \\
  \end{talltblr}
\end{table}


{
For comparison, the last column of Table~\ref{tab:scale_dependence_quadrupole} shows the forecasted $1\sigma$ errors for an ideal cosmic-variance-limited CMB measurement using temperature and $E$-mode polarization up to $\ell=2000$, as obtained in Ref.~\cite{Shiraishi:2016omb}.
Although the marginalized joint constraints obtained here are weaker than the corresponding ideal CMB forecasts for all four values of $\alpha$, the improvement achieved within LSS is more directly assessed by comparison with previous galaxy-clustering measurements. 
}
{
Ref.~\cite{Sugiyama:2017ggb} constrained the quadrupolar power asymmetry
using a BipoSH analysis of BOSS DR12 galaxy clustering without
including the anisotropic-bias response.
For comparison with the present analysis, the published BOSS
constraints are corrected for the anisotropic-bias contribution and
listed in the penultimate column of
Table~\ref{tab:scale_dependence_quadrupole}.
Whereas this correction fixes $b_1^{(2)}$ separately in each redshift bin, the present
analysis treats $b_1^{(2)}$ as a nuisance parameter.
Since the velocity auto-spectrum is independent of $b_1^{(2)}$,
marginalizing over this parameter leaves its constraint on $g_{2M}$
unchanged.
As shown in Table~\ref{tab:scale_dependence_quadrupole}, the
$\mathrm{vv}$-only constraint is slightly tighter than the
corresponding BOSS constraint for $\alpha=-1$ and substantially tighter
for $\alpha=-2$.
Thus, although the ideal CMB forecasts remain stronger, peculiar velocity improves the existing LSS constraints for the red-tilted models without relying on a fixed anisotropic-bias response.
}

The velocity auto-spectrum also provides a consistency check on the inferred primordial power asymmetry when the anisotropic-bias response is modeled incorrectly.  With $\mathrm{gg}$ alone, adopting an incorrect value of $b_1^{(2)}$ changes the inferred $g_{2M}$ because the two contributions enter the same angular sector.  By contrast, the velocity auto-spectrum constrains $g_{2M}$ independently of $b_1^{(2)}$.  Combining it with the density field can therefore reveal such a shift rather than merely reducing the statistical error.  More generally, this robustness applies to uncertainty in the tracer response, while the assumed primordial scale dependence $f_{\rm mod}(k)$ must still be tested separately.


\subsubsection{Constraints from individual power spectra as a function of $k_{\rm max}$}
\label{subsec:kmax_dependence}
{
The preceding forecasts were evaluated at the fixed maximum wavenumber $k_{\rm max}=0.1\,h\,{\rm Mpc}^{-1}$.  To examine how the range of Fourier modes included in the analysis affects the constraints, we now vary $k_{\rm max}$ while fixing $R_N=10$ and $\eta_{\rm cov}=0.7$.  For each value of $k_{\rm max}$, we compute the marginalized error on $g_{2M}$ from each of the three spectra, $\mathrm{gg}$, $\mathrm{gv}$, and $\mathrm{vv}$, and from their joint combination.  We also show the corresponding unmarginalized errors to quantify the degradation caused by the degeneracy with the anisotropic-bias parameter.
}

\begin{figure}[!h]
  \centering
  \begin{minipage}[t]{0.49\textwidth}
  \centering
  \includegraphics[width=\linewidth]{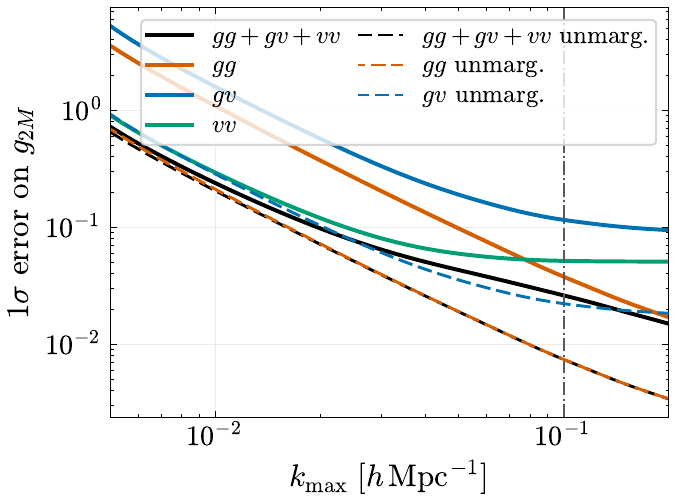}
  \subcaption{$f_{\rm mod}(k)=1$.}
  \end{minipage}
  \hfill
  \begin{minipage}[t]{0.49\textwidth}
  \centering
  \includegraphics[width=\linewidth]{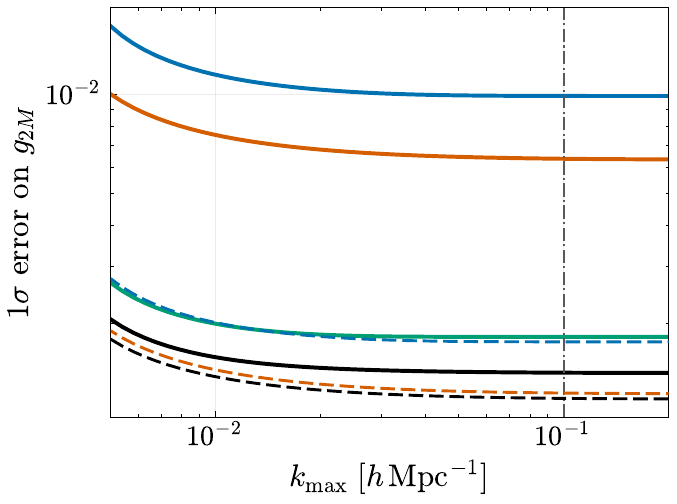}
  \subcaption{$f_{\rm mod}(k)=(k/k_g)^{-2}$.}
  \end{minipage}
  \caption{Marginalized and unmarginalized $1\sigma$ errors on $g_{2M}$ as a function of $k_{\rm max}$, using $R_N=10$ and $\eta_{\rm cov}=0.7$.  The vertical axis shows $\Delta g_{2M}$.  Solid curves include marginalization over the anisotropic-bias parameter, while dashed curves show the corresponding unmarginalized errors.  The dot-dashed vertical lines mark $k_{\rm max}=0.1\,h\,{\rm Mpc}^{-1}$.}
  \label{fig:kmax_core_spectra}
\end{figure}

For the scale-independent case, Fig.~\ref{fig:kmax_core_spectra}(a), the errors continue to decrease as $k_{\rm max}$ is increased.  This is expected because the modulation amplitude is not concentrated on the largest scales, so smaller-scale modes keep adding Fisher information.  At low and intermediate $k_{\rm max}$, the velocity auto-spectrum is competitive despite the fiducial reconstruction noise, reflecting the large-scale velocity response.  At higher $k_{\rm max}$, however, the $\mathrm{vv}$ constraint begins to saturate, while the $\mathrm{gg}$ constraint continues to improve.  The joint analysis remains the most constraining across the whole range.  The separation between the solid and dashed curves shows that marginalization over the anisotropic-bias parameter is a major effect, especially for the density sector.
At $k_{\rm max}=0.1\,h\,{\rm Mpc}^{-1}$, the marginalized errors are $\Delta g_{2M}=2.6\times10^{-2}$, $3.8\times10^{-2}$, $1.2\times10^{-1}$, and $5.2\times10^{-2}$ for the joint, $\mathrm{gg}$, $\mathrm{gv}$, and $\mathrm{vv}$ analyses, respectively.  The corresponding unmarginalized joint error is $7.4\times10^{-3}$.  Thus the joint analysis improves on each individual spectrum, but marginalization over the anisotropic-bias normalization still degrades the joint constraint by a factor of about $3.5$.

Figure~\ref{fig:kmax_core_spectra}(b) shows different behavior for the scale-dependent model.  In this case the signal is dominated by the largest available modes, and the constraints nearly saturate once $k_{\rm max}$ reaches a few times $10^{-2}\,h\,{\rm Mpc}^{-1}$.  Adding smaller-scale modes then gives little additional improvement.  The velocity spectra are particularly useful in this scale-dependent case because the velocity field itself weights large scales more strongly than the density field.  As a result, the joint constraint is set mainly by the complementarity between the density and velocity sectors rather than by the accumulation of high-$k$ modes.
At $k_{\rm max}=0.1\,h\,{\rm Mpc}^{-1}$, the marginalized $\mathrm{gg}$, $\mathrm{gv}$, and $\mathrm{vv}$ errors are $\Delta g_{2M}=6.4\times10^{-3}$, $9.9\times10^{-3}$, and $1.8\times10^{-3}$, respectively, while the joint error is $1.4\times10^{-3}$.  The small difference between the marginalized and unmarginalized joint errors, $1.4\times10^{-3}$ and $1.2\times10^{-3}$, shows that the low-$k$ velocity information has already removed most of the degradation caused by marginalizing over the anisotropic-bias parameter in this large-scale-enhanced case.

\subsubsection{Dependence on velocity-reconstruction noise and coverage}
\label{subsec:noise_dependence}
\begin{figure}[t]
  \centering
  \begin{minipage}[t]{0.49\textwidth}
  \centering
  \includegraphics[width=\linewidth]{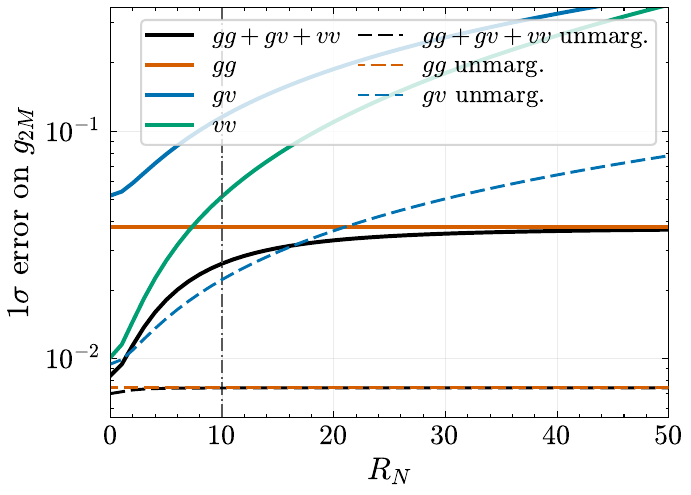}
  \subcaption{Variation of $R_N$ with $f_{\rm mod}(k)=1$.}
  \end{minipage}
  \hfill
  \begin{minipage}[t]{0.49\textwidth}
  \centering
  \includegraphics[width=\linewidth]{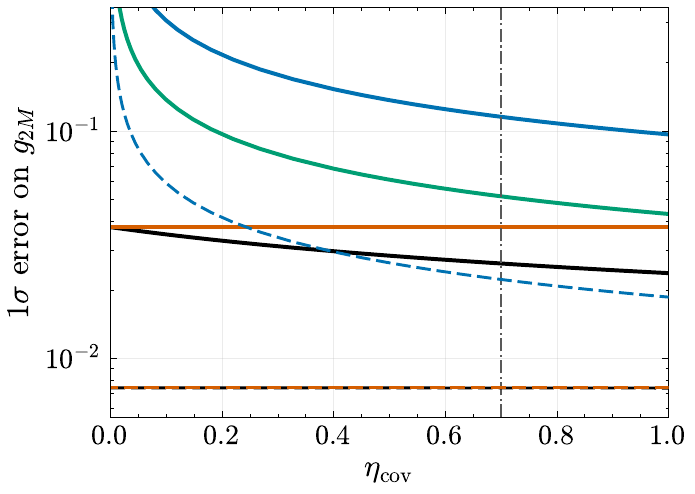}
  \subcaption{Variation of $\eta_{\rm cov}$ with $f_{\rm mod}(k)=1$.}
  \end{minipage}
  \par\medskip
  \begin{minipage}[t]{0.49\textwidth}
  \centering
  \includegraphics[width=\linewidth]{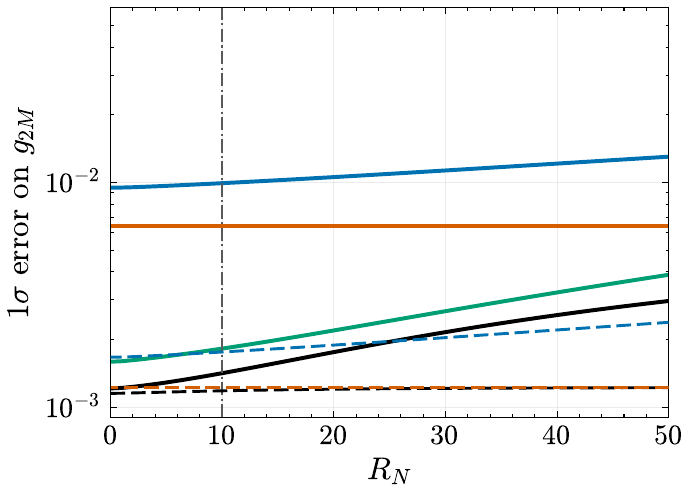}
  \subcaption{Variation of $R_N$ with $f_{\rm mod}(k)=(k/k_g)^{-2}$.}
  \end{minipage}
  \hfill
  \begin{minipage}[t]{0.49\textwidth}
  \centering
  \includegraphics[width=\linewidth]{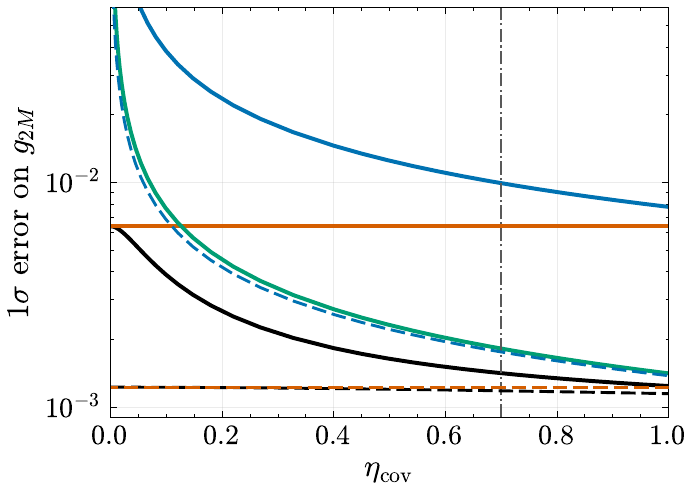}
  \subcaption{Variation of $\eta_{\rm cov}$ with $f_{\rm mod}(k)=(k/k_g)^{-2}$.}
  \end{minipage}
  \caption{Dependence of the $1\sigma$ error on $g_{2M}$ on the velocity-reconstruction noise and the Euclid--CMB overlap fraction for the quadrupolar model.  All panels use $k_{\rm max}=0.1\,h\,{\rm Mpc}^{-1}$.  Solid curves include marginalization over the anisotropic-bias parameter, while dashed curves show the corresponding unmarginalized errors.  In panel (b), the dashed $\mathrm{gg}$ and joint curves are nearly coincident.
  {
  Panels (a) and (c) vary $R_N$ with $\eta_{\rm cov}=0.7$ fixed,
  while panels (b) and (d) vary $\eta_{\rm cov}$ with $R_N=10$
  fixed.  The dot-dashed vertical lines mark $R_N=10$ in panels
  (a) and (c) and $\eta_{\rm cov}=0.7$ in panels (b) and (d).
  }
  Panels (c) and (d) show the corresponding large-scale-enhanced case with $f_{\rm mod}(k)=(k/k_g)^{-2}$ and $k_g=0.05\,{\rm Mpc}^{-1}$.
  }
  \label{fig:rn_coverage_core_spectra}
\end{figure}

{
Having established the forecast for the fiducial values $R_N=10$ and $\eta_{\rm cov}=0.7$, we now examine how the constraints depend on the velocity-reconstruction noise and the CMB--Euclid overlap fraction.  We vary these quantities over $0\leq R_N\leq50$ and $0\leq\eta_{\rm cov}\leq1.0$, respectively, while keeping the galaxy number density, the Euclid footprint, and the fiducial cosmology fixed.  Varying $R_N$ isolates the effect of the reconstruction noise, whereas varying $\eta_{\rm cov}$ isolates the effect of the survey overlap.  The resulting marginalized and unmarginalized errors are shown in Fig.~\ref{fig:rn_coverage_core_spectra}.  We consider both the scale-independent case, $f_{\rm mod}(k)=1$, and the large-scale-enhanced case, $f_{\rm mod}(k)=(k/k_g)^{-2}$ with $k_g=0.05\,{\rm Mpc}^{-1}$.
}

Reducing $R_N$ improves the constraints involving $\mathrm{gv}$ and $\mathrm{vv}$.
The improvement is much more pronounced after marginalizing over $b_1^{(2)}$ than when fixing it because the density--velocity cross-spectrum and velocity auto-spectrum constrain parameter combinations that are complementary to the galaxy auto-spectrum, thereby reducing the $g_{2M}$--$b_1^{(2)}$ degeneracy.
At $R_N\simeq10$, the density spectrum remains important in the scale-independent case, whereas for $\alpha=-2$ the velocity auto-spectrum is already the most constraining individual spectrum, since both $f_{\rm mod}(k)\propto k^{-2}$ and $K_{\rm v}\propto k^{-1}$ enhance the relative contribution of low-$k$ modes. 
Reducing $R_N$ from $10$ to $5$ lowers the marginalized joint error
from $\Delta g_{2M}=2.6\times10^{-2}$ to $1.8\times10^{-2}$ for $\alpha=0$, but only from
$1.4\times10^{-3}$ to $1.3\times10^{-3}$ for $\alpha=-2$.  This weaker dependence on $R_N$ for
$\alpha=-2$ is consistent with the forecast being dominated by 
{
the cosmic variance, i.e., the limited number of Fourier modes at low-$k$.
}
Nevertheless, the small improvement
obtained by reducing $R_N$ below its fiducial value does not imply that
the velocity information is unimportant.  Even when $R_N$ is increased,
the marginalized $\mathrm{vv}$ constraint remains tighter than the
marginalized $\mathrm{gg}$ constraint up to $R_N\simeq50$.  The
marginalized $\mathrm{gv}$ constraint, by contrast, is already weaker
than $\mathrm{gg}$ at $R_N=10$ and deteriorates further as $R_N$
increases.

Increasing $\eta_{\rm cov}$ increases the effective volume over which the density--velocity cross-spectrum and velocity auto-spectrum are available without changing the reconstruction noise per mode.
At fixed $R_N=10$, increasing $\eta_{\rm cov}$ from $0.3$ to $0.7$ reduces the marginalized joint error from $\Delta g_{2M}=3.1\times10^{-2}$ to $2.6\times10^{-2}$ for $\alpha=0$ and from $2.1\times10^{-3}$ to $1.4\times10^{-3}$ for $\alpha=-2$.  Extending the overlap from $0.7$ to unity gives smaller improvements, to $2.4\times10^{-2}$ and $1.2\times10^{-3}$, respectively. 
As $\eta_{\rm cov}$ approaches zero, the result approaches the galaxy-only limit.

\clearpage
\section{Conclusion}
\label{sec:conclusion}

We have developed a unified BipoSH framework for constraining dipolar and quadrupolar power asymmetry with galaxy clustering and line-of-sight peculiar velocity.
These forecasts show that peculiar-velocity power spectra play qualitatively different roles in the dipolar and quadrupolar cases.
The two fields respond to the same underlying matter fluctuations but differ in two essential respects.  Within the linear model adopted here, the galaxy-density field depends on tracer-bias parameters, whereas the peculiar-velocity response is independent of them.  In linear theory, the velocity response contains an additional $k^{-1}$ factor, increasing its relative sensitivity to the largest-scale modes.

For the dipolar asymmetry, 
{the model adopted in this paper introduces no additional contributions to the observables, and there is thus no notable parameter degeneracy with power-asymmetry parameter.}
At the fiducial noise level, the velocity spectra therefore mainly provide consistency checks with observational systematics different from those of galaxy clustering.

{On the other hand, for the quadrupolar power asymmetry,}
galaxy clustering contains both a primordial contribution proportional to $g_{2M}$ and an anisotropic-bias contribution proportional to $g_{2M}b_1^{(2)}$.  The galaxy auto-spectrum alone therefore produces a strongly elongated two-parameter constraint. {Notably,} the density--velocity cross-spectrum has a different degeneracy direction in the $g_{2M}$--$b_1^{(2)}$ plane, and combining it with the lowest-order galaxy BipoSH coefficients provides most of the improvement in the joint constraint. {Furthermore,} the velocity auto-spectrum plays a complementary role.  Because it is independent of $b_1^{(2)}$, it provides a direct constraint on $g_{2M}$ against which the galaxy-based inference can be checked.  For $\alpha=-1$, the $\mathrm{vv}$-only constraint is comparable to, and nominally tighter than, the corresponding BOSS constraint, while for $\alpha=-2$ it is substantially tighter.  Because the velocity auto-spectrum is independent of $b_1^{(2)}$, these constraints are unaffected by marginalization over the anisotropic-bias parameter.  Thus, the cross-spectrum accounts for most of the constraining power gained in the joint analysis, whereas the velocity auto-spectrum provides an independent and competitive constraint on the primordial power asymmetry. Both roles become more effective for the adopted large-scale-enhanced modulation, for which the velocity response and the primordial signal receive greater weight from low-$k$ modes.

Over the ranges considered here, reducing the velocity-reconstruction noise yields a larger fractional improvement for the scale-independent model than for the large-scale-enhanced model, whereas increasing the Euclid--CMB overlap fraction yields a larger fractional improvement for the large-scale-enhanced model.  These forecasts are particularly timely because kSZ velocity reconstruction has progressed from forecasts and simulation tests to three-dimensional measurements, including recent detections of galaxy--velocity and velocity--velocity power spectra with ACT DR6 and DESI DR2~\cite{Deutsch:2017ybc,Smith:2018bpn,Munchmeyer:2018eey,Giri:2020pkk,McCarthy:2024nik,Chaussidon:2026vmn}.

{Finally,} the quantitative accuracy of the forecast {in the present paper} is limited by the use of linear theory, local plane-parallel geometry, Gaussian covariance, independent redshift bins, and a scale-independent velocity-noise model.  
{The three spectra considered in this paper offer complementary degeneracy-breaking power because they respond differently to $b_1^{(2)}$. Nevertheless, their actual quantitative contribution must be re-evaluated after incorporating these systematic effects.}
{A realistic analysis will need to account for several observational and theoretical factors, such as kSZ noise, optical-depth calibration, survey windowing, foreground contamination, reconstruction-induced correlations, and wide-angle, relativistic, or nonlinear corrections}~\cite{Giri:2020pkk,Contreras:2022zdz,Shiraishi:2023zda}.
{Incorporating these systematic effects and validating the pipeline with mock catalogs will be essential steps toward translating our forecast into actual data-level constraints.}

\begin{acknowledgments}
   We are grateful to Toshiki Kurita for his valuable insights and discussions throughout the early stages of this project. 
   This work was supported in part by JST SPRING, Grant Number JPMJSP2110 (KM), JSPS KAKENHI Grant Numbers {23K25868 and 26H02044} (AT), JP20H05859 and JP23K03390 (MS). {AT acknowledges support from FY2025 MUSUBIME of Kyoto University. }
   TO acknowledges support from the Taiwan National Science and Technology Council under Grants Nos. NSTC 112-2112-M-001-034-, NSTC 113-2112-M-001-011-, and NSTC 114-2112-M-001-004-, and the Academia Sinica Investigator Project Grant No. AS-IV-114-M03 for the period of 2025–2029.
MS acknowledges the Center for Computational Astrophysics, National Astronomical Observatory of Japan, for providing the computing resources. 
\end{acknowledgments}

\appendix

\bibliographystyle{JHEP}
\bibliography{references}

\end{document}